# Anomalous transport regimes and asymptotic concentration distributions in the presence of advection and diffusion on a comb structure

Olga A. Dvoretskaya and Peter S. Kondratenko

*Nuclear Safety Institute, Russian Academy of Sciences, 52 Bolshaya Tul'skaya St., 115191 Moscow, Russia*

We study a transport of impurity particles on a comb structure in the presence of advection. The main body concentration and asymptotic concentration distributions are obtained. Seven different transport regimes occur on the comb structure with finite teeth: classical diffusion, advection, quasidiffusion, subdiffusion, slow classical diffusion and two kinds of slow advection. Quasidiffusion deserves special attention. It is characterized by a linear growth of the mean squared displacement. However, quasidiffusion is an anomalous transport regime. We established that a change of transport regimes in time leads to a change of regimes in the space. Concentration tails have a cascade structure, namely consisting of several parts.

## I. INTRODUCTION

Anomalous transport in highly heterogeneous media is a subject of extensive studies for decades [1]. Due to the complexity and variety of real heterogeneous media, general theory of the transport is not yet available for them. For this reason, studying anomalous transport by simple physical models is a matter of exceptional importance. A comb structure is one of such models. This model has much in common with a percolation cluster. A backbone and teeth of the comb structure are like respectively to a backbone and dead (dangling) ends of the percolation cluster. The simplest version of the comb structure based on the classical diffusion equation was analyzed in [2]. A random walk on the comb and comblike structure was studied in Refs. [3], [4]. Subdiffusion with a power $\gamma = 1/4$ was obtained in [2]. Another transport regime has been found and termed as a quasidiffusion in Ref. [5]. This regime is a result of the particles departure into the teeth of the structure and advection. Classical diffusion as a physical transport mechanism for a backbone of the comb structure has been supplemented by longitudinal advection in Refs. [6, 7]. However, the authors couldn't have obtained some interesting results as the backbone and the teeth had an infinite thickness and length in these works. A finite thickness of the comb structure and finite length of the teeth cause additional transport regimes and "power trains". So, the results obtained in Refs. [2-7] do not exhaust all problem aspects of the impurity transport on the comb structure. A pattern of transport regimes is still an open question in the case of finite length teeth, as well as a fine structure of the asymptotic concentration distribution at all time intervals.

The purpose of this paper is a detailed analysis of the impurity transport on the comb structure. In particular, we obtained transport regimes in addition to the known previously and found that the asymptotic concentration distribution often differs from Gaussian form. Also, we established that a transition from one time interval to another may be accompanied by a change of the transport regime in media with a contrast distribution of characteristics. Concentration tails have a cascade structure (see also Refs. [8-10]).

The structure of the paper is as follows: section 2 presents the problem and basic relations; section 3 focuses on obtaining results such as time evolution of the impurities concentration in the backbone. The total number of impurities in the backbone is found in section 4. Main conclusions of the paper are summarized in the final section.

## II. PROBLEM FORMULATION AND BASIC RELATIONS

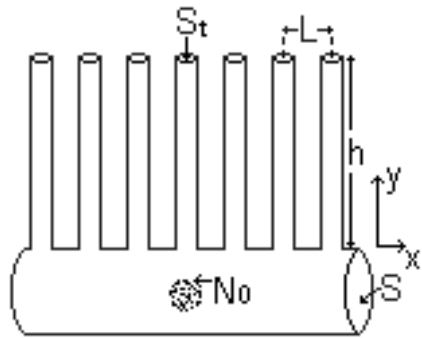

FIG.1. Comb structure

The comb structure is illustrated in Fig.1. It consists of a backbone and a periodic system of teeth. The backbone is a straight cylinder of an infinite length along the $x$ axis. The cross-sectional area of the backbone is equal to S. Each tooth is also a straight cylinder with a length $h$. The boundary between each tooth and the backbone is plane with the size of $S_t$. L is a period of the comb structure.

A transport of impurity particles in the backbone results from advection with a velocity $u$ directed along the $x$ axis and isotropic diffusion with coefficient $D$. The transport occurs by diffusion with coefficient $d$ in each tooth. A concentration of impurity particles is denoted by $n$ in the backbone and $c$ - in a tooth. The impurity particles resided in the backbone are called *active*. A boundary condition on the outer surface of the comb structure is a normal component of zero impurity flux. A condition on the boundary between the backbone and teeth consists in continuity of the concentration and normal component of the flux density. It is assumed that the concentration distribution at the initial moment $t=0$ is specified and located in the region with a size $\sigma_0$ within the backbone.

We will be interested in the concentration distribution at times

$$t \gg \frac{max\{S, L^2\}}{4D} \qquad (1)$$

In this case, there is almost a homogeneous concentration of *active* particles $n(x,t)$ in the backbone. In turn, a concentration inside each tooth depends on its longitudinal coordinate y, time $t$ and coordinate $x$ as a parameter.

Averaging three-dimensional advection-diffusion equation over cross-sectional area and period of the comb structure, we obtain the equation for the concentration of active particles

$$\frac{\partial n(x,t)}{\partial t} + u\frac{\partial n(x,t)}{\partial x} = D\frac{\partial^2 n(x,t)}{\partial x^2} - q(x,t), \qquad (2)$$

where $q(x,t)$ is a magnitude of the impurity flux from the backbone into the teeth per unit of length along the $x$ axis

$$q = -\frac{S_t}{LS} d \frac{\partial c}{\partial y}\bigg|_{y=+0} \qquad (3)$$

The equations for the concentration in the tooth and the corresponding boundary conditions are

$$\frac{\partial c(x,y;t)}{\partial t} = d\frac{\partial^2 c(x,y;t)}{\partial y^2}. \qquad (4)$$

$$\frac{\partial c}{\partial y}\bigg|_{y=h} = 0, \qquad c(0,t;x) = n(x,t). \qquad (5)$$

Now turn to the Fourier-Laplace space $(k, p)$ in Eqs. (2) - (5). Then the solution of Eq. (4) with conditions (5) leads to the following expression for the impurity flux

$$q_{kp} = n_{kp}\sqrt{p/t_1}\, tanh\left(\sqrt{pt_2}\right), \qquad (6)$$

with

$$t_1 = \frac{1}{d}\left(\frac{LS}{S_t}\right)^2, \qquad t_2 = \frac{h^2}{d}. \qquad (7)$$

We further assume that $t_1$ and $t_2$ satisfy the inequality $t_1 \ll t_2$. $t_1$ is the time when the number of particles in the teeth is compared with that in the backbone, while $t_2$ is a characteristic time of diffusion at distances of the order of $h$ into the teeth.

Using Eqs. (2), (6), we find the concentration of active particles in the space $(k, p)$.

$$n_{kp} = \frac{n_{kp}^{(0)}}{p + \sqrt{p/t_1}\, tanh\left(\sqrt{pt_2}\right) + iuk + Dk^2}, \qquad (8)$$

where $n_{kp}^{(0)}$ is a Fourier- Laplace transform of the initial concentration distribution averaged on the cross-sectional area of the backbone $\bar{n}^{(0)}(x) \equiv \bar{n}(x,0)$.

Performing the inverse Fourier-Laplace transformation in Eq. (8) and integrating with respect to $k$, we find

$$n(x,t) = \int_{-\infty}^{+\infty} dx'\, G(x-x',t) n^{(0)}(x), \qquad (9)$$

where Green's function $G(x,t)$ is given by

$$G(x,t) = \frac{1}{u} exp\left(\frac{ux}{D}\theta(-x)\right) \int_{b-i\infty}^{b+i\infty} \frac{dp}{2\pi i} \frac{exp\{-\Phi(p;|x|,t)\}}{\Lambda(p)}; \quad Re\, b > 0 \qquad (10)$$

with

$$\Phi(p;x,t) = \frac{ux}{2D}(\Lambda(p)-1) - pt, \qquad (11)$$

$$\Lambda(p) = \sqrt{1 + t_u \left[p + \sqrt{p/t_1}\, tanh\left(\sqrt{pt_2}\right)\right]} \qquad (12)$$

and $t_u = 4D/u^2$. Evidently, $t_u$ is the time, where a displacement due to advection becomes comparable with a diffusion length.

The transport regime is determined by two important quantities: an average of the displacement $\langle x \rangle$ related to advection and the impurity variance of the displacement $\sigma(t)$

$$\langle x \rangle = \frac{1}{N(t)} \int_{-\infty}^{+\infty} dx\, x\, n(x,t),$$
$$(\sigma(t))^2 = \frac{1}{N(t)} \int_{-\infty}^{+\infty} dx\, (x-\langle x \rangle)^2 n(x,t), \qquad (13)$$

where $N(t) = \int_{-\infty}^{+\infty} dx\, n(x,t)$ is the total number of active particles at time $t$.

Note that the concentration "tails" correspond to this condition: $|x - \langle x \rangle| \gg \sigma(t)$. Further assume that $\sigma(t) \gg \sigma_0$. Thus we have

$$n(x,t) \cong \frac{N_0}{S} G(x,t). \qquad (14)$$

Here the initial total number of the active particles is denoted by $N_0 = S \int_{-\infty}^{+\infty} dx\, \bar{n}^{(0)}(x)$ and the reference point of coordinate $x$ is chosen in the area of the initial concentration distribution. Using Eq. (8), we obtain

$$N(t) = N_0 \int_{-\infty}^{+\infty} dx\, G(x,t) \tag{15}$$

Hereafter, we consider $G(x,t)$ at positive $x$. In order to find the expression for Green's function at negative $x$, one can use Eq. (10).

## III. TRANSPORT REGIMES AND ASYMPTOTIC CONCENTRATION DISTRIBUTION

The Green function behavior and asymptotic concentration structure depend on a relation between characteristic times $t_u$, $t_1$ and $t_2$. Let us analyze the problem separately for each of these relation and characteristic time intervals.

We stress that the main body concentration is determined by $pt \sim 1$ in Eq. (10). In turn, the concentration tails correspond to $pt > 1$.

**1.** $t_u \ll t_1$

**1.1.** $t \ll t_u$

This case is obtained as a limit $u \to 0$, $t_1 \to \infty$. Therefore Green's function is given by

$$G(x,t) = (4\pi Dt)^{-1/2} \exp\left(-\frac{x^2}{4Dt}\right) \tag{16}$$

This is a well known classical diffusion expression.

**1.2.** $t_u \ll t \ll t_h$

We have $t_2^{-1} \ll p \ll t_u^{-1}$ for the main body concentration. Therefore, we make use of the following expressions for $\Lambda(p)$ and $\Phi(p;x,t)$ of Eqs. (11), (12)

$$\Lambda(p) \cong 1 + \frac{t_u}{2}\left(p + \sqrt{\frac{p}{t_1}} - \frac{1}{4}t_u p^2\right), \tag{17}$$

$$\Phi(p;x,t) = \frac{x}{u}\left(\sqrt{\frac{p}{t_1}} - \frac{1}{4}t_u p^2\right) - pt', \tag{18}$$

with $t' = t - \frac{x}{u}$ (recall that $x > 0$).

We show below that the $G$-function behavior essentially depends on whether a current time is more or less than the characteristic time $t_3 = \left(t_u t_1^2\right)^{1/3}$. Formally, this is determined by which of the two terms in parentheses of Eq. (18) is a dominant under integration in Eq. (10). Let us analyze the cases $t_u \ll t \ll t_3$ and $t_3 \ll t \ll t_2$ separately.

**1.2a.** $t_u \ll t \ll t_3$

First we consider the main body concentration as a dependence on the spatial variable $x$. We suppose that significant values of $p$ in Eq. (10) are determined by the term in the exponent from Eq. (18) at $x \simeq ut$. Then we have $p \sim (t_u t)^{-1/2}$. At these values of $p$, the term in Eq. (18) proportional to $\sim \sqrt{p}$ is estimated as $\frac{x}{u}\sqrt{\frac{p}{t_1}} \sim \left(\frac{x}{ut_3}\right)^{3/4}$. Further calculations reveal that a spatial variable $x$ satisfies the inequality $x < ut$ in the main body concentration. Thus we obtain the strong inequality

$$\frac{x}{u}\sqrt{\frac{p}{t_1}} < \left(\frac{t}{t_3}\right)^{3/4} \ll 1. \tag{19}$$

It confirms the assumption made above relative to predominance of term $\sim p^2$, hence allowing us to neglect the term $\sim \sqrt{p}$ in Eq. (18) while calculating the integral in Eq. (10). As a result, we get

$$G(x,t) \cong (4\pi Dt)^{-1/2} \exp\left\{-\frac{(x-ut)^2}{4Dt}\right\}. \tag{20}$$

This expression corresponds to the classical advection. Here the average of the displacement and the impurity variance of the displacement are $\langle x \rangle = ut, \quad \sigma = \sqrt{2Dt}$ and so $\sigma \ll \langle x \rangle$ .

The Eq. (20) is valid at the distances not too far from the peak of $G$-function. To evaluate Green's function at the large distances, we take advantage of saddle-point technique while integrating in Eq. (10) with respect to $p$. The saddle-point is given by equation $\frac{\partial}{\partial p_0}\Phi(p_0;x,t)=0$ and takes the value $p_0 = -t'/2t_u t$. Note that $p_0$ has a real value and a sign, opposite to a sign of $t'$. Expression (20) remains valid in the right wing of $G$-function (i.e. where $x > ut$), because the original contour of integration encounters no singularities in Eq. (10) while shifting toward the saddle-point. Another situation occurs in the left wing (where $t' > 0$.). Here the saddle-point is negative. Therefore shifting the integration contour to saddle-point, we meet the branch point $p = 0$ of the integrand in Eq. (10). It results from terms $\sim \sqrt{p}$ in $\Phi(p;x,t)$ and $\Lambda(p)$. Hence we should take into account a contribution $\delta_b G$ from the integration along the banks of the cut from the branch point. To find above contribution, we can neglect the term $\sim p^2$ in $\Phi(p;x,t)$. Then we substitute Eqs. (17), (28) in Eq. (10) and expand to the first order. Finally, we find

$$\delta_b G(x,t) \cong \frac{ut_u + 2x}{4u^2} \frac{1}{\sqrt{\pi t_1 t'^3}} \quad (21)$$

This contribution due to the unusual behavior of the concentration distribution at times $t_u \ll t \ll t_3$. Possessing a power decrease $\propto (x-ut)^{-3/2}$ the contribution $\delta_b G$ has advantage over exponentially decreasing expression (20) at the relatively far distances from the peak. By comparison of Eq. (20) and Eq. (21), we conclude that power contribution (21) dominates at the condition $t' > \sqrt{t_u t \ln\left(\frac{t_3}{t}\right)}$.

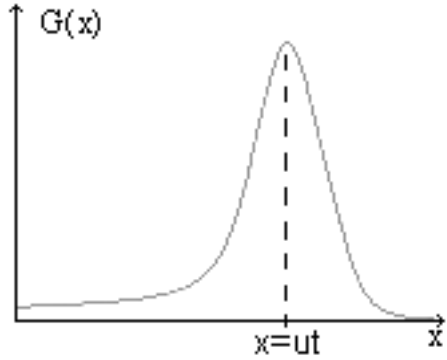

FIG. 2. Qualitative behavior of Green's function at times $t_u \ll t \ll t_3$

At times $t_u \ll t \ll t_3$ we have a regime similar to the classical advection with nearly symmetrical shape of the concentration distribution, slightly "spoiled" by the presence of a power train [see Eq. (21)]. That behavior of Green's function is illustrated in Fig.2

**1.2b.** $t_3 \ll t \ll t_2$

In accordance with the estimate (19), terms $\sim p^2$ and $\sqrt{p}$ in Eq. (18) exchange roles under a transition from the interval $t_u \ll t \ll t_3$ to the interval $t_3 \ll t \ll t_1$. Therefore, we get the following approximation

$$\Phi(p;x't') \cong \frac{x}{u}\sqrt{\frac{p}{t_1}} - pt'. \quad (22)$$

A substitution of Eq. (22) in Eq. (10) gives the expression

$$G(x,t') = \frac{x + ut_u/2}{ut'} \frac{1}{\sqrt{4\pi D_u t'}} exp\left(-\frac{x^2}{4D_u t'}\right), \quad (23)$$

with $D_u = u^2 t_1$.

A similar behavior of the concentration distribution (to our knowledge) has not been found before.

To find asymptotic concentration profiles we take advantage of saddle-point technique. There are two saddle-point when $t' > 0$. Clearly, it is worth it to take into account such a saddle-point that leads to a smaller exponent value. It follows that such saddle-point is $p_0 = \frac{x^2}{4D_u t'^2}$ at times $t' \gg t(t_u/2t_1)^{1/3}$. This contribution is reduced to expression (23).

Since only one saddle-point $p_0 \cong (t_u^2 t_1)^{-1/3}$ remains in the region $|t'| \ll t(t_u/2t_1)^{1/3}$, we have $G$-function for this value of $p_0$

$$G(x,t) \propto exp\left(-\frac{3t}{4t_3}\right) \quad (24)$$

Also one saddle-point $p_0 = -t'/2t_u t$ takes place in the case $t' < 0$ and $|t'| \gg t(t_u/2t_1)^{1/3}$. Hence this contribution is determined by Eq. (20).

Expression (23) applies to the whole time interval $t_3 \ll t \ll t_2$. However there is fundamental difference between case $t_3 \ll t \ll t_1$ and $t_1 \ll t \ll t_2$.

At times $t_3 \ll t \ll t_1$ the average of the displacement is $\langle x \rangle = ut$, and the impurity variance of the displacement (width of the peak) is $\sigma \sim ut^2/t_1$. Obviously, in that case $\langle x \rangle \gg \sigma$. Consequently one can replace the numerator in exponent of Eq. (23) by $(ut)^2$. Thus we have

$$G(x,t') = \frac{x + \frac{ut_u}{2}}{ut'} \frac{1}{\sqrt{4\pi D_u t'}} exp\left(-\frac{t^2}{4t_1 t'}\right). \quad (25)$$

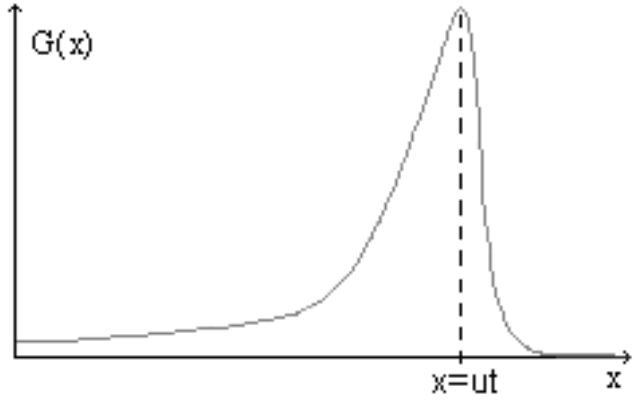

FIG. 3. Qualitative behavior of Green's function at times $t_3 \ll t \ll t_1$

In that way, we have advection with sharply asymmetric spatial concentration distribution at $t_3 \ll t \ll t_1$. This is illustrated in Fig.3. Namely the left wing of the concentration distribution is characterized by a power low and the right wing corresponds to a rapid exponential decay of Eq. (25) followed by Gaussian decrease.

The presence of power trains is unexpected in the cases where $t \ll t_1$, because prima facie teeth cannot yet become significant at these times.

However, such unusual behavior of the concentration distribution has a physical meaning and qualitative explanation. The peak of distribution had reached the tooth and some of active particles had been going into the tooth. Then the peak of the distribution moved behind the tooth and particles came back to the backbone. Thus the presence of power trains is due to departure of active particles into the teeth and subsequent comeback into the backbone.

At times $t_1 \ll t \ll t_2$ the average of the displacement and the impurity variance of the displacement have the same order $\langle x \rangle \sim \sigma \sim \sqrt{D_u t}$ and $\sqrt{D_u t} \ll ut$. Hence one can replace $t'$ by $t$ applying the main body concentration and the first stage of the tail in Eq. (23). Finally, we find

$$G(x,t) = \frac{x + \frac{ut_u}{2}}{ut} \frac{1}{\sqrt{4\pi D_u t}} exp\left(-\frac{x^2}{4D_u t}\right). \tag{26}$$

Similar transport regime has been found in [5] and termed as quasidiffusion. The second and third stages of the tail coincide with the first and the second stage in case 1.2.b. Notice that $\sigma \sim t^{1/2}$ takes place in quasidiffusion similar to classical diffusion but the total number of active particles is not retained. So quasidiffusion is an anomalous transport regime.

**1.3.** $t \gg t_2$

At these times the most significant values of Laplace variable are $p \ll t_2^{-1}$ for the main body concentration and the first stage of tail. Thus we have approximation $\tanh(\sqrt{pt_2}) \cong \sqrt{pt_2} - \frac{1}{3}(\sqrt{pt_2})^3$. Using Eqs. (11) and (12), we get

$$\Phi(p;x,t) \cong -\frac{x\tilde{D}_u}{\tilde{u}^3}p^2 - p\tilde{t}', \qquad pt_2 \ll 1, \qquad (27)$$

here we denote $\tilde{u} = u\sqrt{\frac{t_1}{t_2}}$, $\tilde{D}_u \cong \frac{1}{3}D_u$, $\tilde{t}' = t - \frac{x}{\tilde{u}}$.

Substituting Eq. (27) into Eq. (10), we obtain

$$G(x,t) \cong \frac{1}{\sqrt{4\pi\tilde{D}_u t}}\sqrt{\frac{t_1}{t_2}}\exp\left\{-\frac{(x-\tilde{u}t)^2}{4\tilde{D}_u t}\right\}. \qquad (28)$$

This is classical advection with modified advection velocity and modified diffusion coefficient. So we called it slow advection. The average of the displacement $\langle x \rangle = \tilde{u}t$ and the impurity variance of the displacement $\sigma \simeq \sqrt{2\tilde{D}t}$. Hence, $\sigma \ll \langle x \rangle$.

Along with the main body concentration Eq. (29) describes also the active particles distribution at the first stage of the tail until the saddle-point satisfies the inequality $p_0 t_2 < 1$. The second stage of the tail is determined by $p_0 \gtrsim t_2^{-1}$ and given by Eq. (26). It corresponds to quasidiffusion. An approximate border between tail stages meets the distance $x - \tilde{u}t \sim \tilde{u}t$ and $G$-function is

$$G(x,t) \propto \exp(-t/t_2). \qquad (29)$$

The third stage begins from distances $|t'| \sim t(t_u/2t_1)^{1/3}$ and has a form (20), where $|t'| > t(t_u/2t_1)^{1/3}$

**2.** $t_1 \ll t_u^2/t_1 \ll t_2$

While calculating Green's function at times smaller then $t_u^2/t_1$ one can use an approximation

$$\Lambda(p) \cong \frac{2}{u}\sqrt{D\left(p + \sqrt{\frac{p}{t_1}}\right)}, \qquad t \ll t_u. \qquad (30)$$

**2.1.** $t \ll t_1$

This case is entirely similar to the **1.1**.

**2.2.** $t_1 \ll t \ll t_u^2/t_1$

The term $\sim p$ should be neglected under the root. Combining Eq. (30) and Eq. (10), we obtain

$$G(x,t) \cong \frac{1}{2}\left(\frac{t_1}{D^2 t^3}\right)^{1/4} F(\xi), \quad \xi = \frac{x}{\sqrt{D\sqrt{t_1 t}}}.$$

$$F(\xi) = \int_{a-i\infty}^{a+i\infty} \frac{ds}{2\pi i} s^{-1/4} \exp\left(\xi s^{1/4} - s\right), \quad s = pt, \quad \operatorname{Re} a > 0.$$

(31)

This expression corresponding to subdiffusion was found early in Refs. [2, 6, 7, 10, 11]. The impurity variance of the displacement has an estimation $\sigma \sim \sqrt{D\sqrt{t_1 t}}$. Using Eq. (31), we get the first stage of asymptotic Green's function

$$G(x,t) \cong \frac{1}{2\sqrt{6}}\left(\frac{t_1}{D^2 t^3}\right)^{1/4} \left(\frac{\xi}{4}\right)^{1/3} \exp\left\{-3\left(\frac{\xi}{4}\right)^{4/3}\right\}.$$

(32)

It was also found in [2]. The second stage of the tail ($p_0 t_1 > 1$) corresponds to the classical diffusion expression (16). $G(x,t) \sim (4\pi D t)^{-1/2} \exp(-t/t_1)$ at sample boundary between the second and the first stage of the tail (where $\xi \sim 4(t/3t_1)^{3/4}$).

**2.3.** $t_u^2/t_1 \ll t \ll t_2$

In this time interval quasidiffusion expression (26) holds for the main body concentration and the first stage of the tail. The second stage of the tail corresponds to Eq. (32) and the third stage is described by classical diffusion [see Eq. (16)].

**2.4.** $t \gg t_2$

Here the deduction formally coincides with the case of 1.3 for the main body concentration and first stage of the tail, leading to the expression for the slow advection [see Eq. (28)]. At these times the tail consists of four stages. The second, third and fourth stages are determined by Eq. (23), (20) and Eq. (16) respectively.

A detailed analysis showed that the "tails" have a cascade structure and the following regularity takes place: with increasing distances such a transport regime occurs what was realized in the main body of concentration at an earlier time interval. Earlier these properties of the tails were established in Refs. [8-10]. In the next case a tails consideration is omitted, because the above-mentioned regularity also is valid.

**3.** $t_u^2/t_1 \gg t_2$

**3.1.** $t \ll t_1$ corresponds to case **1.1.**

**3.2.** $t_1 \ll t \ll t_2$ is entirely similar to the **2.2.**

### 3.3. $t_2 \ll t \ll t_u\sqrt{\frac{t_2}{t_1}}$

In this case, one can use approximation $tanh(\sqrt{pt_2}) \cong \sqrt{pt_2}$, $u \to 0$ to the main body concentration and first stage of tail. It now follows that

$$G(x,t) \cong \frac{1}{\sqrt{4\pi \tilde{D} t}} \sqrt{\frac{t_1}{t_2}} exp\left\{-\frac{x^2}{4\tilde{D}t}\right\}, \qquad (33)$$

where $\tilde{D} = \sqrt{t_1/t_2}\, D$.

This expression corresponds to the slow classical diffusion.

### 3.4. $t \gg t_u\sqrt{\frac{t_2}{t_1}}$

This case formally is similar to the 1.3. So, the Green function takes a form (28) for the main body concentration and the first stage of the tail. But the effective diffusion coefficient should be replaced by $\tilde{D} = \sqrt{t_1/t_2}\, D$.

## IV. TOTAL NUMBER OF ACTIVE PARTICLES

In order to find the total number of active particles $N(t)$, we take advantage of obvious relations

$$N(t) = \int_{b-i\infty}^{b+i\infty} \frac{dp}{2\pi i} n_{kp}\Big|_{k=0}, \qquad N(0) \equiv N_0 = n_k^{(0)}\Big|_{k=0}, \qquad (34)$$

where $n_{kp}$ is defined by Eq. (8).

$N(t)$ is given by simple expressions in three cases

$$\begin{aligned} N(t) &\cong N_0, & t \ll t_1; \\ N(t) &\cong N_0 \sqrt{\frac{t_1}{\pi t}}, & t_1 \ll t \ll t_2; \\ N(t) &\cong N_0 \sqrt{\frac{t_1}{t_2}}, & t \gg t_2. \end{aligned} \qquad (35)$$

This means that at times $t \ll t_1$ the relative number of particles into the teeth has been very small yet. Therefore, the total number of active particles almost coincides with its initial value $N_0$. In the case $t \gg t_1$, most of the impurity particles are located in the teeth and a ratio $N(t)/N_0$ is inversely proportional to the volume of the teeth occupied by impurity particles. At times $t_1 \ll t \ll t_2$ particles go into the teeth very intensively and $N(t) \sim t^{-1/2}$. A similar notation was made in Ref. [11]. The relation $N(t) \sim t^{-1/2}$ also was found in Refs. [2, 4, 8, 11, 12]. Finally at $t \gg t_2$ the teeth are saturated with the impurity particles, and again become stationary but $N(t) \ll N_0$. We see that the teeth of the comb structure act as traps. A similar effect occurs in the percolation media [10, 13]

## V. CONCLUSION

In the present work we have studied in detail the transport of impurity particles on a comb structure in the presence of advection and diffusion in the backbone. All obtained results are easily generalized to a random comb structure. Also, our results are valid for a random statistically homogeneous comb structure.

We have obtained a main body concentration and a concentration distribution at the large distances (concentration tails). Seven different transport regimes are realized. Each regime is determined by the relation between characteristic times and a considered time interval. Thus the following transport regimes occur: classical diffusion, subdiffusion, slow classical diffusion, quasidiffusion, classical advection and two kind of slow advection. The first three regimes exist due to the presence of diffusion, moreover the second and the third significantly result from the departure of impurity particles from the backbone into the teeth. The next four regimes are caused by "interaction" of advection and the particles departure into the teeth. Three additional regimes (two kinds of slow advection and slow classical diffusion) arise on the comb structure with finite teeth compared with the structure of infinite teeth.

The impurity transport in the presence of diffusion only was studied in Refs. [2], [11] where the authors found typical transport regimes – classical diffusion and subdiffusion for the comb structure with infinite teeth. Besides, following notation was also developed in Ref. [2]: a finite length of teeth results in an additional regime – slow classical diffusion. Various modifications of the comb structure were considered in [11]. It should be noted that transport regimes arising due to the presence of advection have not been studied as well as a fine structure of concentration tails in above mentioned works.

Our analysis showed that the concentration tails have a cascade structure in all transport regimes except for classical diffusion. The results confirmed the regularity, which earlier was established in Refs. [8-10]: with increasing distances such a transport regime occurs what was realized in the main body of concentration at an earlier time interval. Thus the change of transport regimes occurs in both time and space.

Some characteristics of advection seem to be unexpected at times $t_u \ll t \ll t_1$. In this time interval, the number of particles located in the teeth is still relatively small. However the influence of the particles departure into the teeth determines to considerable extent the spatial width of the concentration distribution peak. Also this phenomenon results in a power law decrease of the concentration distribution in the left wing. A faster decrease than Gaussians occurs, namely $G(x,t) \sim \exp\left(-t^2 / 4t_1(t - x/u)\right)$ in the right wing.

It should be noted that many authors defined anomalous diffusion as diffusion with a nonlinear growth of the mean squared displacement [14-18]. That definition is not full. For example, quasidiffusion is an anomalous transport regime because the total number of active particles is not conserved, although the variance of the displacement in this regime depends on time as $\sigma \propto t^{1/2}$ just as in classical diffusion.

## ACKNOWLEDGMENT

This work supported by the Russian Foundation of Basic Research (RFBR) under project 08-08-01009a.